\documentclass[aps,prb,twocolumn,superscriptaddress]{revtex4-2}

\usepackage{graphicx}
\usepackage{mathrsfs}
\usepackage{bm}
\usepackage{amsmath}
\usepackage{color}
\usepackage{dcolumn}
\usepackage{epstopdf}
\usepackage{dsfont}
\usepackage{amssymb}
\usepackage{tabularx}
\usepackage{array,ulem}
\usepackage{float}
\usepackage{colordvi}
\usepackage{verbatim}
\usepackage{hyperref}

\hypersetup{
colorlinks=true,
linkcolor=blue,
filecolor=blue,
urlcolor=blue,
citecolor=blue,
}

\begin{document}
\title{
Altermagnetism-induced higher-order topological Dirac semimetals in two-dimensional nonsymmorphic systems
}

\author{Lizhou Liu}
\affiliation{International Center for Quantum Materials, School of Physics, Peking University, Beijing 100871, China}

\author{Cheng-Ming Miao}
\affiliation{International Center for Quantum Materials, School of Physics, Peking University, Beijing 100871, China}

\author{Ying-Tao Zhang}
\affiliation{College of Physics, Hebei Normal University, Shijiazhuang 050024, China}

\author{Qing-Feng Sun}
\email[Correspondence author:~~]{sunqf@pku.edu.cn}
\affiliation{International Center for Quantum Materials, School of Physics, Peking University, Beijing 100871, China}
\affiliation{Hefei National Laboratory, Hefei 230088, China}

\date{\today}

\begin{abstract}
We propose an altermagnetism-induced route to a two-dimensional higher-order topological Dirac semimetal in a trilayer nonsymmorphic Dirac system.
Interlayer hybridization preserves the nonsymmorphic bulk Dirac crossings and generates helical edge states in the trilayer phase.
Introducing outer-layer $d$-wave altermagnetic order gaps the helical edge states, while only the symmetry-protected Dirac crossing at high-symmetry point $M$ remains gapless.
The resulting phase hosts a surviving bulk Dirac point, gapped edge spectra, and corner states.
An exact layer-space decomposition further reveals that these features arise from the coexistence of a gapless Dirac sector and a gapped $\mathbb{Z}_2$ topological sector.
Our results establish altermagnetism as an effective mechanism for engineering higher-order topological Dirac semimetals in two-dimensional nonsymmorphic systems.
\end{abstract}

\maketitle

\section{Introduction}

Topological semimetals, characterized by symmetry-protected band crossings near the Fermi level, provide a fertile platform for unconventional topological responses~\cite{Haldane2004, Hasan2010, Qi2011, Wan2011, Xu2011, Armitage2018, Lv2021, Liu2025dw}.
Among them, Dirac semimetals are particularly important because their fourfold-degenerate Dirac points can be viewed as superposed Weyl points with opposite chirality, making them natural parent states for symmetry-breaking-induced topological transitions~\cite{Young2012, Wang2012, Wang2013, Liu2024ds}.
While Dirac semimetals are well established in three dimensions~\cite{Wang2012, Wang2013, Liu2014Na3Bi, Liu2014Cd3As2, Borisenko2014, Neupane2014, He2014}, their robust two-dimensional counterparts are much more elusive, since spin-orbit coupling generally gaps Dirac crossings \cite{KaneMele2005a}.
A major advance was made by Young and Kane, who showed that nonsymmorphic symmetries can protect two-dimensional Dirac points even in the presence of spin-orbit coupling~\cite{Young2015}.

Another major development is higher-order topological states~\cite{Benalcazar2017, Benalcazar2017a, Langbehn2017, Schindler2018, Ren2020, Liu2024, Mandal2025, Liu2025a, Miao2025, Miao2025a}, which extend the bulk-boundary correspondence beyond conventional edge or surface states.
In such phases, a two-dimensional bulk can host zero-dimensional corner states~\cite{Liu2024Braiding, Miao2024, Liu2024hotd}, while a three-dimensional bulk can host one-dimensional hinge states~\cite{Wang2021}.
This concept has been extended to gapless systems, leading to higher-order topological semimetals in which bulk band degeneracies coexist with lower-dimensional boundary modes~\cite{Lin2018, Ghorashi2020, Wang2020, Wieder2020, Li2020, Qiu2021, Wang2023Floquet, Qi2024}.
In three-dimensional Dirac semimetals~\cite{Lin2018, Szabo2020, Ghorashi2021, Wang2022HODSM, Zeng2023, Pal2025}, bulk Dirac points can separate momentum slices with distinct higher-order topology, giving rise to hinge states terminating at the projected nodes.
The two-dimensional case is more restrictive because zero-dimensional corner states must coexist with symmetry-protected bulk Dirac points. Such coexistence is difficult to achieve, since boundary mass terms that gap edge modes often also gap or shift the bulk Dirac crossings.
A viable mechanism must therefore gap the boundary spectrum while preserving the bulk Dirac point.

Altermagnetism provides a natural route toward this goal~\cite{Smejkal2022b, Smejkal2022a, McClarty2024}.
As a compensated collinear magnetic order, it combines vanishing net magnetization with symmetry-enforced momentum-dependent spin splitting. Unlike a conventional Zeeman field, this spin splitting is sign-changing and strongly anisotropic, with a momentum-space structure dictated by crystal symmetry and realizable in different symmetry channels~\cite{Smejkal2022a, Roig2024, Liu2024Twisted, Liu2026Odd}.
Recent theoretical advances have established altermagnetism as a platform for unconventional transport responses~\cite{Yi2026Spin, Cheng2024JJ, Sun2025MR, Yu2026Magnon, Wang2026Photon, Dong2025Photocurrent} and diverse topological states~\cite{Parshukov2025, Sun2025TSC, Ghorashi2024, GonzalezHernandez2025SCN, Liu2026cpb, Yi2026Majorana, Liu2026Chirality, Wan2025Arc, Zhu2024Dislocation}. These features naturally raise the question of whether altermagnetism can provide the missing ingredient for realizing higher-order topology in a two-dimensional Dirac semimetal.

\begin{figure}
  \centering
  \includegraphics[width=8.5 cm,angle=0]{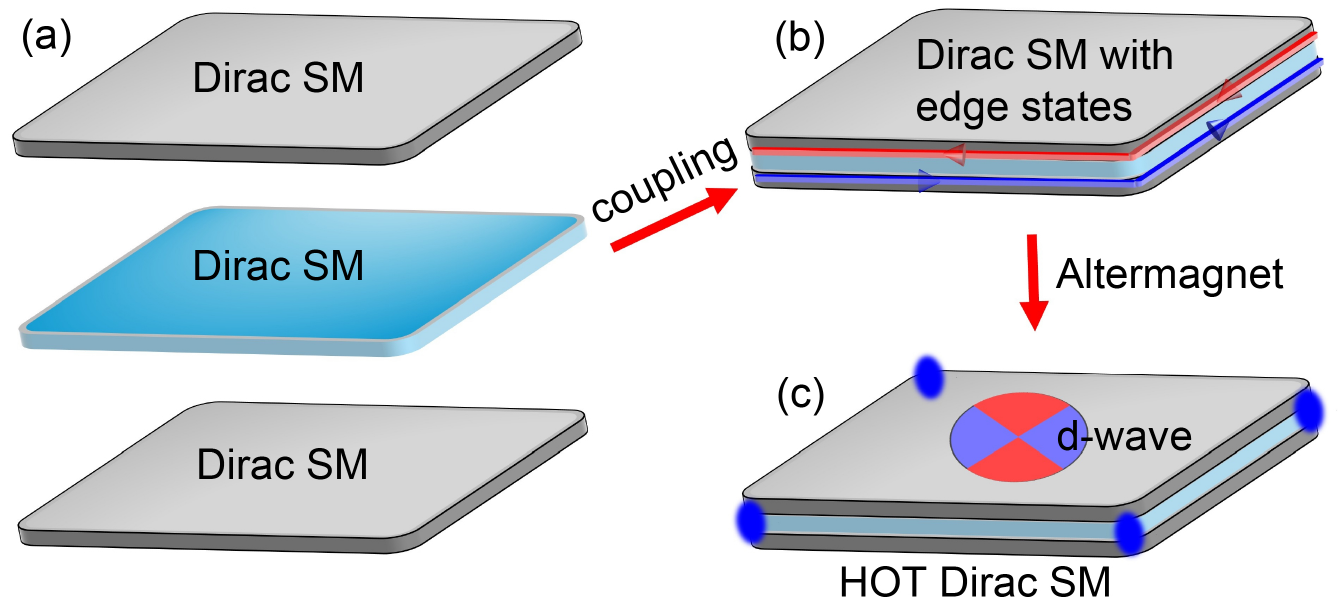}
\caption{
(a) Three decoupled two-dimensional nonsymmorphic Dirac semimetal (SM) layers.
(b) Interlayer coupling preserves the bulk Dirac semimetal character and generates boundary states, shown by the red and blue edge channels.
(c) Trilayer system with $d$-wave altermagnetic order on the two outer gray layers. The blue dots denote corner states.
The resulting phase is labeled as a higher-order topological Dirac semimetal (HOT Dirac SM).
}
  \label{fig1}
\end{figure}

In this work, we answer this question by proposing a route to an altermagnetism-induced higher-order topological Dirac semimetal based on a trilayer nonsymmorphic Dirac semimetal.
Starting from a trilayer extension of a two-dimensional nonsymmorphic Dirac semimetal as shown in Fig.~\ref{fig1}(a)~\cite{Young2015}, we show that interlayer hybridization separates the system into a gapless Dirac sector and a gapped topological sector supporting helical edge states [Fig.~\ref{fig1}(b)].
An outer-layer $d$-wave altermagnetic order gaps the helical edge states by inducing opposite Dirac masses on adjacent edges, while preserving only the symmetry-protected Dirac crossing at the high-symmetry point $M$ [Fig.~\ref{fig1}(c)].
The resulting phase hosts a surviving bulk Dirac point, gapped edge spectra, and corner states, thereby realizing a two-dimensional higher-order topological Dirac semimetal.

\section{System Hamiltonian}

We construct a trilayer extension of the nonsymmorphic Dirac model as shown in Fig.~\ref{fig1}(a)~\cite{Young2015}, where each layer contains two sublattices and spin degrees of freedom.
We use the basis $\Psi_{\mathbf{k}}=(\psi_{1,\mathbf{k}},\psi_{2,\mathbf{k}},\psi_{3,\mathbf{k}})^T$, with $\psi_{\ell,\mathbf{k}}=(c_{\ell A\uparrow}(\mathbf{k}),c_{\ell A\downarrow}(\mathbf{k}),c_{\ell B\uparrow}(\mathbf{k}),c_{\ell B\downarrow}(\mathbf{k}))^T$ and $\ell=1,2,3$ labeling the layer index.
Here $\tau_i$ and $s_i$ denote Pauli matrices acting on the sublattice $(A, B)$ and spin $(\uparrow,\downarrow)$ degrees of freedom, respectively.
For compactness, we define
\begin{align}
h_t(\mathbf{k})
&=
2t\cos\frac{k_x}{2}\cos\frac{k_y}{2}\,\tau_x s_0,
\nonumber\\
g(\mathbf{k})
&=
t_{\mathrm{so}}\tau_z
\left(
s_y\sin k_x-s_x\sin k_y
\right).
\end{align}
The first outer layer and the middle layer are then described by
\begin{equation}
H_1(\mathbf{k})=h_t(\mathbf{k})-g(\mathbf{k}),
\qquad
H_2(\mathbf{k})=h_t(\mathbf{k})+g(\mathbf{k}),
\label{eq:H12}
\end{equation}
while the third layer is taken to be identical to the first one,
$H_3(\mathbf{k})=H_1(\mathbf{k})$.
The first term describes nearest-neighbor intralayer hopping, and the second term is the spin-orbit coupling. The two outer layers, therefore, share the same spin-orbit chirality, whereas the middle layer carries the opposite one.
With a spin- and sublattice-independent hopping $t_{\perp}$ between adjacent layers [Fig.~\ref{fig1}(b)], the trilayer Hamiltonian reads
\begin{equation}
H_{0}(\mathbf{k})
=
\begin{pmatrix}
H_1(\mathbf{k}) & t_{\perp}\,\tau_0 s_0 & 0 \\
t_{\perp}\,\tau_0 s_0 & H_2(\mathbf{k}) & t_{\perp}\,\tau_0 s_0 \\
0 & t_{\perp}\,\tau_0 s_0 & H_3(\mathbf{k})
\end{pmatrix}.
\label{eq:H0}
\end{equation}
Here the off-diagonal blocks describe the coupling between adjacent layers.

We further introduce a $d$-wave altermagnetic term on the two outer layers [Fig.~\ref{fig1}(c)],
\begin{equation}
h_{\mathrm{m}}(\mathbf{k})
=
J(\cos k_x-\cos k_y)\tau_z s_z,
\label{eq:hAM}
\end{equation}
where $J$ denotes the strength of the altermagnetic order. The form factor $\cos k_x-\cos k_y$ represents a $d$-wave realization of the symmetry-dictated sign-changing spin splitting characteristic of altermagnetism~\cite{Smejkal2022b,Smejkal2022a,McClarty2024}. In the trilayer space, the corresponding contribution reads
\begin{equation}
H_{\mathrm{AM}}(\mathbf{k})
=
\begin{pmatrix}
h_{\mathrm{m}}(\mathbf{k}) & 0 & 0 \\
0 & 0 & 0 \\
0 & 0 & h_{\mathrm{m}}(\mathbf{k})
\end{pmatrix},
\label{eq:HAM}
\end{equation}
so that the altermagnetic order acts only on the two outer layers and leaves the middle layer unaffected.
The full Bloch Hamiltonian studied below is
\begin{equation}
H_{3\mathrm{L}}(\mathbf{k})
=
H_{0}(\mathbf{k})+H_{\mathrm{AM}}(\mathbf{k}).
\label{eq:H3L}
\end{equation}
This Hamiltonian defines the trilayer model used in the following analysis.

\begin{figure}
  \centering
  \includegraphics[width=8.5 cm,angle=0]{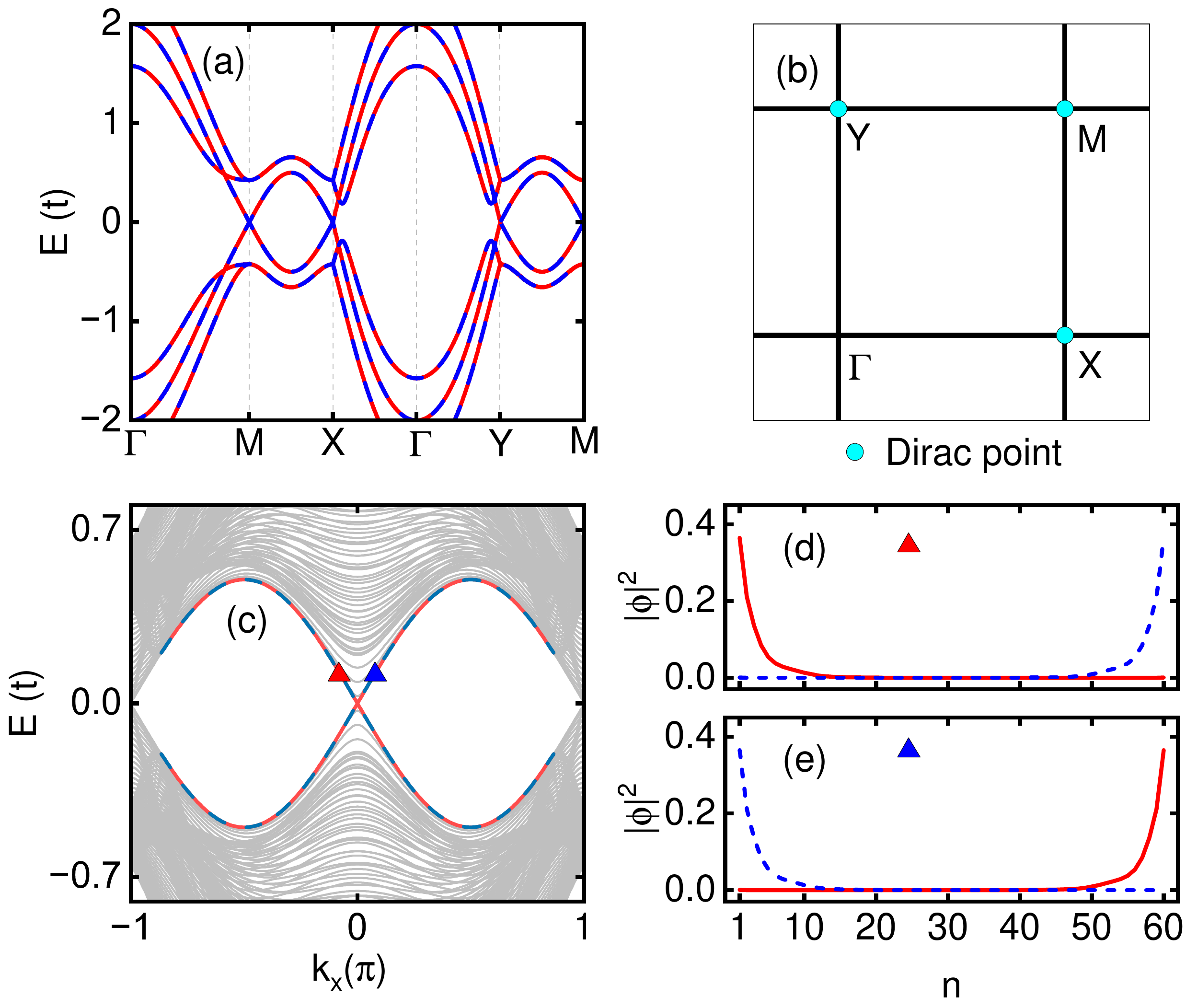}
\caption{
(a) Bulk band structure along $\Gamma$-$M$-$X$-$\Gamma$-$Y$-$M$.
(b) Brillouin zone, where the cyan dots mark the Dirac points at $X=(\pi,0)$, $Y=(0,\pi)$, and $M=(\pi,\pi)$.
(c) Ribbon spectrum with periodic boundary conditions along $x$ and open boundary conditions along $y$.
The red and blue triangles mark two representative edge states.
(d),(e) Probability distributions of the two marked states in (c) as functions of the site index $n$.
Parameters are $t=-1$, $t_{\mathrm{so}}=0.5$, $t_{\perp}=0.3$, and $J=0$.
}
\label{fig2}
\end{figure}

\section{Altermagnetism-induced higher-order topological Dirac semimetal}

We first examine the parent trilayer Hamiltonian $H_0(\mathbf{k})$ in the absence of the $d$-wave altermagnetic term.
The bulk band structure along the high-symmetry path $\Gamma$-$M$-$X$-$\Gamma$-$Y$-$M$ is shown in Fig.~\ref{fig2}(a), while the corresponding Brillouin zone is displayed in Fig.~\ref{fig2}(b).
Three symmetry-protected Dirac crossings are located at the $X$, $Y$, and $M$ points.
Thus, before the altermagnetic term is introduced, the trilayer system realizes a two-dimensional nonsymmorphic Dirac phase with multiple bulk Dirac points.

We next examine the boundary spectrum of the parent phase.
Figure~\ref{fig2}(c) shows the ribbon spectrum obtained with periodic boundary conditions along $x$ and open boundary conditions along $y$.
In addition to the projected bulk bands, a pair of gapless edge branches crosses near $k_x=0$.
The red and blue triangles mark two representative states on these branches.
Their probability distributions are plotted in Figs.~\ref{fig2}(d) and \ref{fig2}(e), respectively.
The red-marked state is localized predominantly near one boundary, whereas the blue-marked state is localized near the opposite boundary.
These results confirm that the crossing branches are helical edge states of the parent trilayer phase.
Therefore, the $J=0$ system simultaneously hosts bulk Dirac crossings and gapless one-dimensional boundary states.

We now introduce the $d$-wave altermagnetic term $H_{\mathrm{AM}}(\mathbf{k})$.
As shown in Fig.~\ref{fig3}(a), a finite $J$ opens a gap at the crossing point of the helical edge states.
The red and blue branches remain localized at the ribbon boundaries but no longer cross at zero energy, indicating that the altermagnetic term acts as a mass term for the parent helical edge states~\cite{Liu2024HOTM}.
To determine whether the gapped boundary spectrum gives rise to higher-order boundary states, we further calculate the energy spectrum of a square nanoflake with open boundaries along both $x$ and $y$.
As shown in Fig.~\ref{fig3}(b), four states appear at zero energy, marked by the blue dots, while representative low-energy bulk states closest to zero energy are marked by the red dots.
The lower inset shows that the four zero-energy states are localized at the four corners of the nanoflake, 
whereas the upper inset shows that the nearby bulk states are spatially extended over the sample.
The four zero-energy states are therefore identified as corner states by their zero-dimensional localization after the helical edge spectrum is gapped.

The formation of these corner states is not accompanied by a complete bulk gap opening.
As shown in Fig.~\ref{fig3}(c), the Dirac crossings at $X$ and $Y$ are gapped after the $d$-wave altermagnetic term is introduced, whereas the crossing at $M$ remains intact at the Fermi level.
The corresponding Brillouin zone is shown in Fig.~\ref{fig3}(d), where the surviving Dirac point at $M$ is marked by the cyan dot.
Therefore, the altermagnetic term simultaneously gaps the helical edge states and selectively removes the bulk Dirac crossings at $X$ and $Y$, while preserving the gapless Dirac excitation at $M$.
The resulting phase thus retains a gapless bulk together with gapped edges and corner-localized states.

The resulting phase therefore exhibits three characteristic features simultaneously: a gapless bulk Dirac point at $M$, gapped one-dimensional edge states, and localized zero-dimensional corner states.
This combination identifies the system as a higher-order topological Dirac semimetal.
The microscopic origin of the selective bulk gap opening and the sign-changing boundary masses will be analyzed below through the exact layer-space block decomposition.

\section{Layer-space block decomposition}
To elucidate the microscopic origin of the bulk and boundary features, we analyze the trilayer Hamiltonian in layer space. Owing to the identical outer layers and symmetric interlayer coupling, the system admits an exact decomposition into two decoupled sectors.

\begin{figure}
  \centering
  \includegraphics[width=8.7 cm,angle=0]{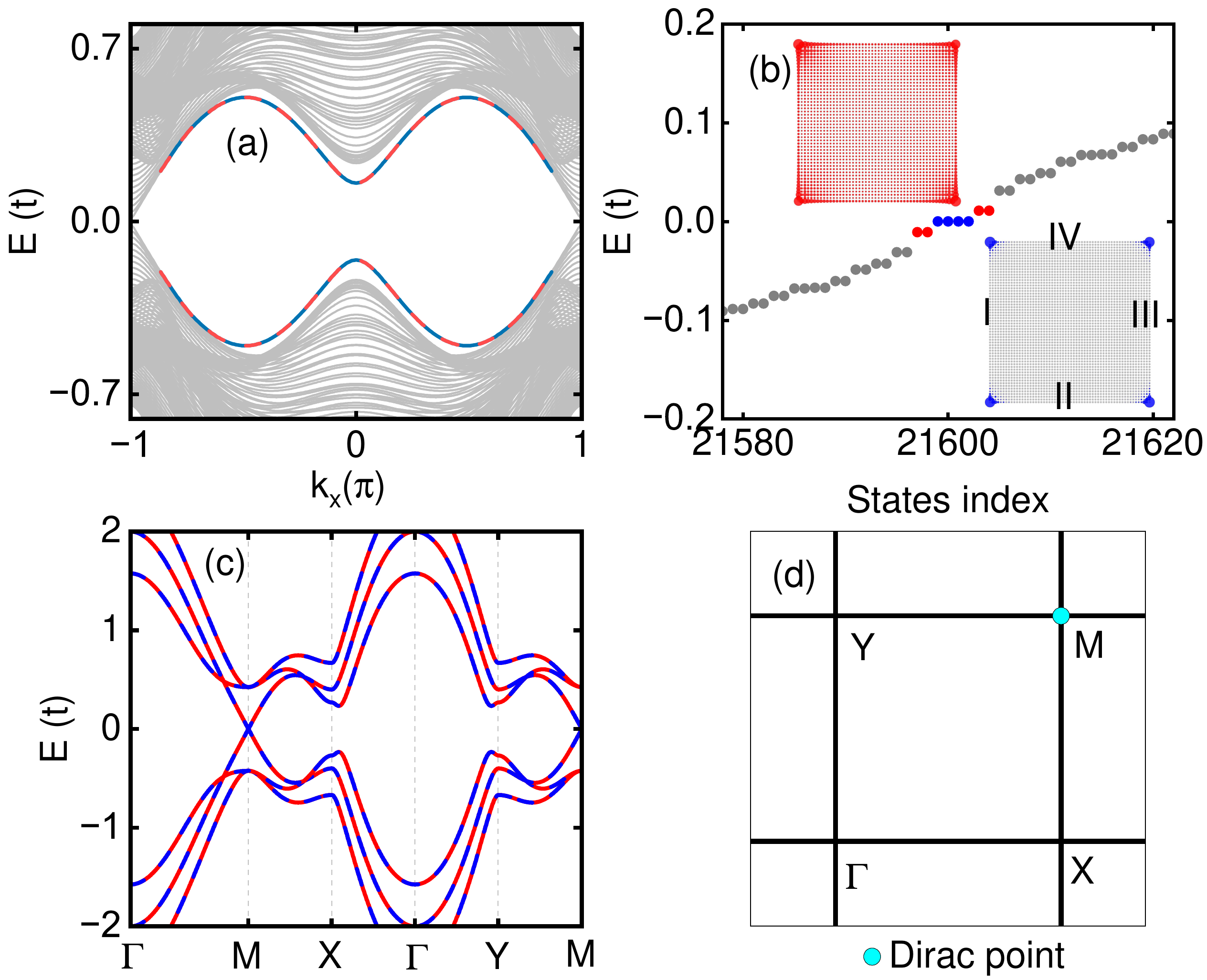}
\caption{
(a) Ribbon spectrum as a function of $k_x$. The red and blue curves denote the gapped edge modes.
(b) Energy levels near zero energy for a square nanoflake. 
The blue dots denote the zero-energy corner states, while the red dots denote representative low-energy states closest to zero energy. The lower and upper insets show the corresponding real-space probability distributions of the blue and red states, respectively.
The Roman numerals  $I$, $II$, $III$, and $IV$ in the lower inset label the four edges of the square nanoflake.
(c) Bulk band structure along $\Gamma$-$M$-$X$-$\Gamma$-$Y$-$M$.
(d) Brillouin zone with the surviving Dirac point at $M$ marked by the cyan dot.
Parameters are $t=-1$, $t_{\mathrm{so}}=0.5$, $t_{\perp}=0.3$, and $J=0.2$.
}
  \label{fig3}
\end{figure}

In layer space, we denote the three layer basis states by $|1\rangle$, $|2\rangle$, and $|3\rangle$, corresponding to the upper outer layer, the middle layer, and the lower outer layer, respectively.
Introducing antisymmetric and symmetric combinations of the two outer-layer basis states,
\begin{equation}
|a\rangle=\frac{|1\rangle-|3\rangle}{\sqrt{2}},\qquad
|s\rangle=\frac{|1\rangle+|3\rangle}{\sqrt{2}},
\end{equation}
the trilayer Hamiltonian can be exactly block-diagonalized as
\begin{equation}
\widetilde{H}_{3\mathrm{L}}(\mathbf{k})
=
H_a(\mathbf{k})\oplus H_{2s}(\mathbf{k}).
\label{eq:block_main}
\end{equation}
The antisymmetric block
\begin{equation}
H_a(\mathbf{k})=h_t(\mathbf{k})-g(\mathbf{k})+h_{\mathrm{m}}(\mathbf{k})
\end{equation}
is completely decoupled from interlayer hybridization, while the hybridized block
\begin{equation}
H_{2s}(\mathbf{k})
=
h_t(\mathbf{k})\eta_0
+
g(\mathbf{k})\eta_z
+
\sqrt{2}t_{\perp}\tau_0s_0\eta_x
+
\frac{1}{2}h_{\mathrm{m}}(\mathbf{k})(\eta_0-\eta_z)
\end{equation}
describes the coupled subspace formed by the middle layer and the symmetric outer-layer channel.
Here $\eta_i$ are Pauli matrices acting on the $(|2\rangle,|s\rangle)$ subspace.
{The detailed derivation of Eq.~\eqref{eq:block_main} is provided in Appendix~\ref{app:block_decomposition}.}
This exact decomposition holds for both $J=0$ and finite $J$.
It separates the problem into a Dirac block $H_a$ and a hybridized block $H_{2s}$, which will be analyzed separately below.
The two blocks also have natural counterparts in systems with different layer numbers: $H_a$ has the same form as the corresponding monolayer Hamiltonian, whereas $H_{2s}$ has an effective bilayer structure in the $(|2\rangle,|s\rangle)$ subspace.

We first consider the trilayer system at $J=0$, where the Hamiltonian separates into an antisymmetric Dirac block $H_a^{(0)}$ and a hybridized block $H_{2s}^{(0)}$ as shown in Eq.~\eqref{eq:block_main}.
The antisymmetric block $H_a^{(0)}(\mathbf{k})=h_t(\mathbf{k})-g(\mathbf{k})$ is completely decoupled from the middle layer and is equivalent to a Young-Kane type two-dimensional nonsymmorphic Dirac Hamiltonian~\cite{Young2015}.
It therefore inherits the nonsymmorphic-symmetry-protected Dirac crossings at $X$, $Y$, and $M$, accounting for the bulk Dirac points in Fig.~\ref{fig2}.

By contrast, the hybridized block $H_{2s}^{(0)}$ is fully gapped and topologically nontrivial.
Setting $J=0$ gives
\begin{equation}
H_{2s}^{(0)}(\mathbf{k})
=
h_t(\mathbf{k})\eta_0
+
g(\mathbf{k})\eta_z
+
\sqrt{2}t_{\perp}\tau_0s_0\eta_x,
\label{eq:H2s0_main}
\end{equation}
where $\eta_i$ are Pauli matrices acting on the $(|2\rangle,|s\rangle)$ subspace.
Defining
\begin{equation}
A(\mathbf{k})
=
2t\cos\frac{k_x}{2}\cos\frac{k_y}{2},
\end{equation}
the corresponding bulk spectrum is
\begin{equation}
E_{\chi,\lambda}(\mathbf{k})
=
\chi
\sqrt{
t_{\mathrm{so}}^2
\left(
\sin^2 k_x+\sin^2 k_y
\right)
+
\left[
A(\mathbf{k})
+
\lambda\sqrt{2}t_{\perp}
\right]^2},
\label{eq:H2s_spectrum_main}
\end{equation}
where $\chi=\pm1$ labels the conduction and valence branches, and $\lambda=\pm1$ labels the two hybridized branches.
Each branch remains twofold degenerate because the combined $\mathcal{P}_{2s}\mathcal{T}$ symmetry leaves $\mathbf{k}$ invariant and satisfies
$(\mathcal{P}_{2s}\mathcal{T})^2=-1$, with
\begin{equation}
\mathcal{T}=is_y\mathcal{K},
\qquad
\mathcal{P}_{2s}=\tau_xs_0\eta_0.
\end{equation}

A bulk gap closing requires both terms under the square root in Eq.~\eqref{eq:H2s_spectrum_main} to vanish.
At the four time-reversal-invariant momenta,
\begin{equation}
A(\Gamma)=2t,
\qquad
A(X)=A(Y)=A(M)=0.
\end{equation}
Therefore, the spectrum at $X$, $Y$, and $M$ is gapped for $t_{\perp}\neq0$, whereas the gap at $\Gamma$ closes only when
\begin{equation}
|t_{\perp}|=\sqrt{2}|t|.
\end{equation}
Thus, the hybridized block is fully gapped in the parameter regime
\begin{equation}
0<|t_{\perp}|<\sqrt{2}|t|.
\label{eq:H2s_gapregion_main}
\end{equation}
Since $H_{2s}^{(0)}$ preserves both time-reversal and inversion symmetries, its $\mathbb{Z}_2$ topology can be determined using the Fu-Kane parity criterion~\cite{FuKane2007}.
At the time-reversal-invariant momenta, the spin-orbit term vanishes, and the occupied eigenstates can be chosen as parity eigenstates of
$\mathcal{P}_{2s}$.
For the fully gapped regime in Eq.~\eqref{eq:H2s_gapregion_main}, the parity products of the occupied Kramers pairs are
\begin{equation}
\delta_{\Gamma}=+1,
\qquad
\delta_X=\delta_Y=\delta_M=-1.
\label{eq:parity_products_main}
\end{equation}
The Fu-Kane invariant is therefore
\begin{equation}
(-1)^{\nu}
=
\delta_{\Gamma}\delta_X\delta_Y\delta_M
=
-1,
\end{equation}
which gives
$
\nu=1.
$
Hence, $H_{2s}^{(0)}$ realizes a two-dimensional quantum spin Hall block and accounts for the helical edge states observed in Fig.~\ref{fig2}(c).
The parent trilayer phase is therefore block resolved: the antisymmetric block $H_a^{(0)}$ produces the bulk Dirac points at $X$, $Y$, and $M$, whereas the hybridized block $H_{2s}^{(0)}$ produces the helical edge states.

We now consider the finite $J$ case.
Since the two outer layers remain identical, the layer-space decomposition remains exact, and the altermagnetic term affects the two sectors in distinct ways.
For the antisymmetric block $H_a$, the parent Young-Kane Dirac phase at $J=0$ is protected by inversion $\mathcal P$, time-reversal symmetry $\mathcal T$, the glide symmetry $\widetilde M_z$, and the two screw symmetries $\widetilde C_{2x}$ and $\widetilde C_{2y}$~\cite{Young2015}.
The scalar form factor of the $d$-wave altermagnetic term,
$
f_d(\mathbf{k})=\cos k_x-\cos k_y,
$
changes sign under the fourfold rotation $C_{4z}$, and therefore acts differently on the three Dirac points of the parent phase.
The altermagnetic term breaks $\mathcal P$, $\mathcal T$, $\widetilde M_z$, and $C_{4z}$ separately, while preserving the combined symmetry $\mathcal{PT}$, the two screw symmetries $\widetilde C_{2x}$ and $\widetilde C_{2y}$, and the magnetic symmetry $C_{4z}\mathcal T$.
Since $C_{4z}$ exchanges the $X$ and $Y$ points, $f_d(\mathbf{k})$ takes finite values with opposite signs at these two momenta, generating constant Dirac masses.
At $X$ and $Y$, the parent protection associated with $\mathcal P$, $\mathcal T$, and $\widetilde M_z$ is therefore removed, while the residual $C_{4z}\mathcal T$ symmetry relates the two opposite masses rather than forcing them to vanish.
By contrast, the $M$ point is invariant under $C_{4z}$ up to a reciprocal lattice vector, and remains invariant under the residual magnetic little-group symmetries. The preserved screw symmetries $\widetilde C_{2x}$ and $\widetilde C_{2y}$, together with $\mathcal{PT}$ and $C_{4z}\mathcal T$, forbid a momentum-independent Dirac mass at $M$. Consequently, the surviving crossing is symmetry protected rather than accidental. Consistently, the sign-changing form factor vanishes at $M$,
$
f_d(M)=0,
$
so that the altermagnetic contribution begins only at higher order in momentum.
{The corresponding low-energy expansions are given in Appendix~\ref{sec:finiteJ_Dirac}.}
As a result, the altermagnetic term gaps the Dirac crossings at $X$ and $Y$, while the Dirac crossing at $M$, protected by nonsymmorphic symmetry, remains gapless.


By contrast, the hybridized block $H_{2s}$ already forms a gapped topological sector at $J=0$ and hosts helical edge states.
Once the altermagnetic term is introduced, these edge states acquire a finite Dirac mass.
The projected perturbation is proportional to $\cos k_x-\cos k_y$, implying an orientation-dependent boundary mass.
Projecting onto the helical edge subspace yields an effective massive Dirac Hamiltonian whose mass sign is fixed by the boundary orientation, so that adjacent edges acquire masses with opposite signs {(see Appendix~\ref{sec:finiteJ_boundary})}.
This sign-alternating edge mass generates domain walls at the corners, which bind localized zero-dimensional states via the Jackiw-Rebbi mechanism~\cite{JackiwRebbi1976}.

The higher-order topology of the gapped $H_{2s}$ sector can be further characterized by mirror-graded winding numbers defined along the two diagonal mirror-invariant lines $k_x=\pm k_y$, as derived in Appendix~\ref{app:mirror_winding}.
For the parameters considered here, both diagonal lines carry nonzero mirror-graded winding numbers with $|\nu_{\mathcal{M}_{\pm}}|=1$, providing an independent bulk characterization of the higher-order topology of the $H_{2s}$ sector.

This block-resolved picture also clarifies the nature of the low-energy states surrounding the zero-energy corner modes in Fig.~\ref{fig3}(b).
The zero-energy corner states originate from the gapped $H_{2s}$ sector, whereas the nearby low-energy bulk states originate from the gapless Dirac sector $H_a$.
They are therefore spatially extended bulk states rather than additional corner modes, consistent with the real-space distributions shown in 
the upper inset of Fig.~\ref{fig3}(b).
Thus, the trilayer combines within a single system the gapless Dirac physics represented by the monolayer-like $H_a$ sector and the higher-order boundary physics represented by the bilayer-like $H_{2s}$ sector.

\begin{figure}
  \centering
  \includegraphics[width=8.7 cm,angle=0]{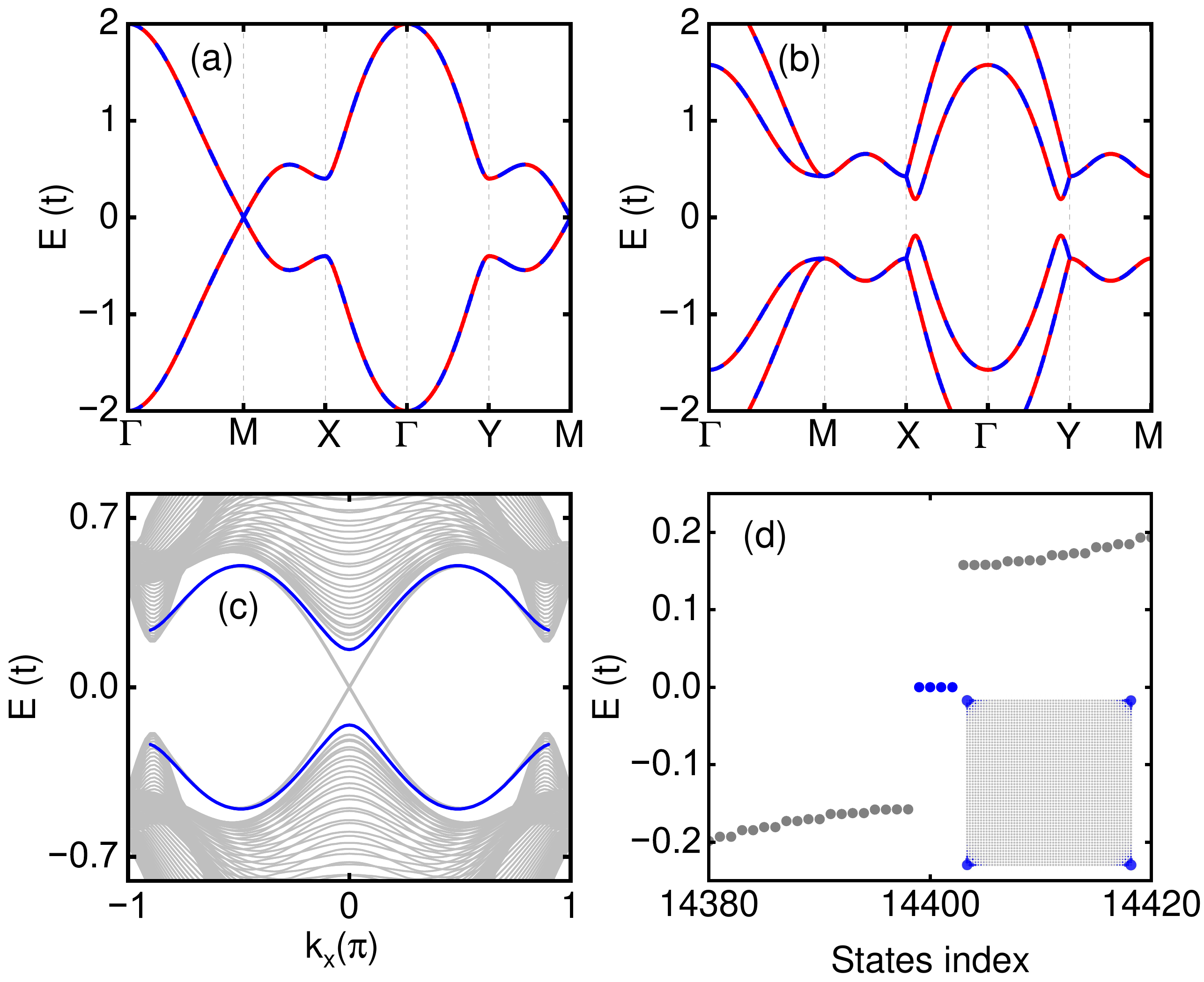}
\caption{
(a) Bulk band structure of the monolayer-like $H_a$ sector with altermagnetic order.
(b) Bulk band structure of the bilayer-like $H_{2s}$ sector with finite interlayer coupling.
(c) Nanoribbon spectra of the $H_{2s}$ sector without and with altermagnetic order. The gapless edge states without altermagnetic order are shown in gray, while the gapped edge states with altermagnetic order are highlighted in blue.
(d) Energy spectrum of the finite $H_{2s}$ system with altermagnetic order. The inset shows the real-space probability distribution of the zero-energy states.
The parameters are the same as those used in Fig.~\ref{fig3}.
}
\label{fig:layer_comparison}
\end{figure}

To further illustrate the physical roles of the two sectors, we show the corresponding results for the monolayer-like $H_a$ sector and the bilayer-like $H_{2s}$ sector in Fig.~\ref{fig:layer_comparison}.
For the $H_a$ sector [Fig.~\ref{fig:layer_comparison}(a)], the altermagnetic term gaps the Dirac crossings at $X$ and $Y$, while the Dirac crossing at $M$ remains gapless.
For the $H_{2s}$ sector, interlayer hybridization opens a bulk gap [Fig.~\ref{fig:layer_comparison}(b)].
The corresponding nanoribbon hosts gapless helical edge states at $J=0$, which become gapped after the altermagnetic term is introduced [Fig.~\ref{fig:layer_comparison}(c)].
The finite system further exhibits four zero-energy corner states localized at the corners [Fig.~\ref{fig:layer_comparison}(d)].
Thus, the $H_a$ sector provides the gapless Dirac component, whereas the $H_{2s}$ sector provides the higher-order topological component with gapped edges and corner states.
Together with the surviving Dirac crossing at $M$, this establishes a higher-order topological Dirac semimetal with a bulk Dirac point and corner states, consistent with Fig.~\ref{fig3}.

\section{Robustness}

We first examine the effect of extending the $d$-wave altermagnetic order to the middle layer.
We introduce the middle-layer altermagnetic term
\begin{equation}
h_{\mathrm{m}}^{\mathrm{mid}}(\mathbf{k})
=
J_m(\cos k_x-\cos k_y)\tau_zs_z,
\end{equation}
where $J_m$ denotes the altermagnetic strength on the middle layer.

Because the two outer layers remain identical, the exact layer-space decomposition in Eq.~\eqref{eq:block_main} remains valid.
Moreover, the antisymmetric state
\begin{equation}
|a\rangle
=
\frac{|1\rangle-|3\rangle}{\sqrt{2}}
\end{equation}
has no weight on the middle layer.
Therefore, the antisymmetric Dirac block remains
\begin{equation}
H_a(\mathbf{k})
=
h_t(\mathbf{k})
-
g(\mathbf{k})
+
h_{\mathrm{m}}(\mathbf{k}),
\end{equation}
and is independent of $J_m$.

By contrast, the hybridized block becomes
\begin{align}
H_{2s}(\mathbf{k})
={}&
h_t(\mathbf{k})\eta_0
+
g(\mathbf{k})\eta_z
+
\sqrt{2}t_{\perp}\tau_0s_0\eta_x
\nonumber\\
&+
\frac{1}{2}
h_{\mathrm{m}}(\mathbf{k})
(\eta_0-\eta_z)
+
\frac{1}{2}
h_{\mathrm{m}}^{\mathrm{mid}}(\mathbf{k})
(\eta_0+\eta_z).
\end{align}
Thus, the middle-layer altermagnetic order modifies only the hybridized block $H_{2s}$ and does not directly affect the surviving Dirac block $H_a$.

\begin{figure*}
\centering
\includegraphics[width=0.9 \textwidth]{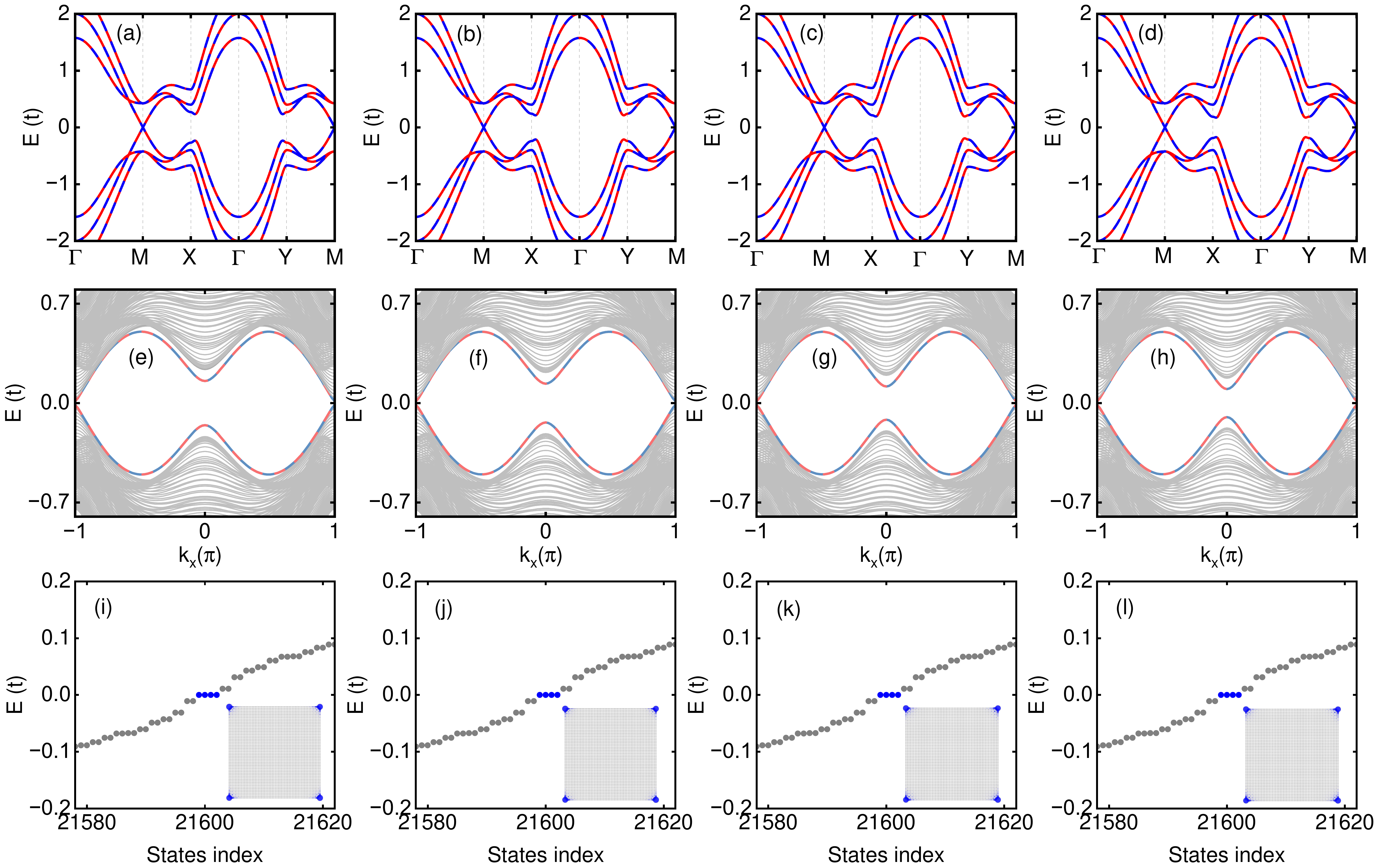}
\caption{
(a)--(d) Bulk band structures for $J_m=0$, $0.1J$, $0.2J$, and $0.3J$, respectively, where $J_m$ denotes the strength of the $d$-wave altermagnetic order introduced on the middle layer.
(e)--(h) Corresponding nanoribbon spectra for the same values of $J_m$ as in (a)-(d).
(i)--(l) Corresponding finite-nanoflake spectra near zero energy.
The blue dots in (i)--(l) denote the four states closest to zero energy, and the insets show their corresponding real-space probability distributions.
The other parameters are the same as those used in Fig.~\ref{fig3}.
}
\label{fig:middle_AM}
\end{figure*}

The corresponding numerical results are shown in Fig.~\ref{fig:middle_AM}.
Figures~\ref{fig:middle_AM}(a)--\ref{fig:middle_AM}(d) show that the bulk Dirac crossing remains gapless as $J_m$ increases from $0$ to $0.3J$.
The nanoribbon spectra in Figs.~\ref{fig:middle_AM}(e)--\ref{fig:middle_AM}(h) show that the boundary gap gradually decreases but remains finite over the parameter range considered.
Correspondingly, the four zero-energy corner states persist in the finite-nanoflake spectra shown in Figs.~\ref{fig:middle_AM}(i)--\ref{fig:middle_AM}(l).

Therefore, extending the same $d$-wave altermagnetic order to the middle layer does not drive the system into a higher-order topological insulator.
The surviving bulk Dirac crossing remains intact because the middle-layer altermagnetic term does not enter the antisymmetric block $H_a$.
These results also show that the proposed higher-order topological Dirac semimetal does not require the altermagnetic order to be perfectly confined to the two outer layers.

\begin{figure*}
\centering
\includegraphics[width=\textwidth]{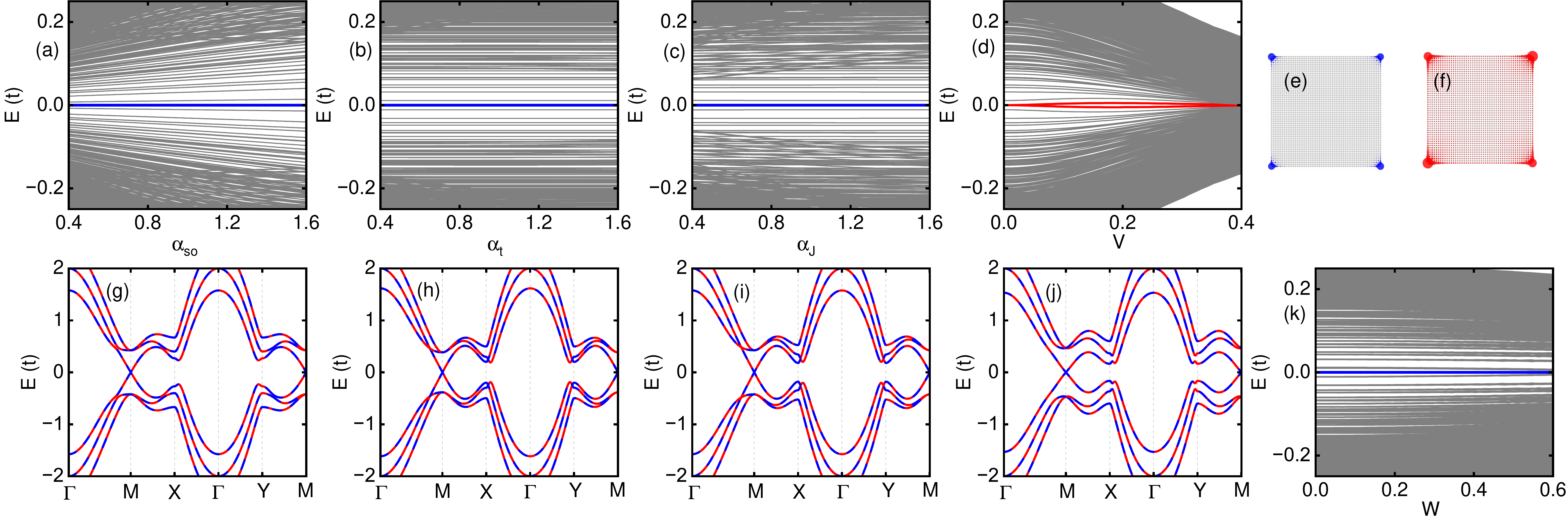}
\caption{
(a)--(d) Finite-flake spectra as functions of $\alpha_{\mathrm{so}}$, $\alpha_t$, $\alpha_J$, and $V$, respectively.
The blue lines in (a)--(c) highlight the four zero-energy corner states, while the red lines in (d) indicate the four corner-derived low-energy states.
(e) Representative real-space probability distribution of the zero-energy corner states for the asymmetric cases in (a)--(c), with the corresponding asymmetry parameter set to $0.8$.
(f) Real-space probability distribution of the four states closest to zero energy for $V=0.2$.
(g)--(j) Bulk band structures for $\alpha_{\mathrm{so}}=0.8$, $\alpha_t=0.8$, $\alpha_J=0.8$, and $V=0.2$, respectively.
(k) Finite-flake spectrum as a function of the onsite magnetic-disorder strength $W$.
The blue lines in (k) highlight the four corner states near zero energy.
The other parameters are the same as those used in Fig.~\ref{fig3}.
}
\label{fig4}
\end{figure*}

Having established the robustness against a finite extension of the altermagnetic order to the middle layer, we next consider more general perturbations that modify the equivalence between the two outer layers.
Unlike the middle-layer altermagnetic term discussed above, these perturbations generally modify or break the exact layer-space decomposition and therefore provide a more stringent test of whether the higher-order topological Dirac semimetal relies on fine tuning.
Specifically, we consider asymmetries in the outer-layer spin-orbit coupling, interlayer coupling, and altermagnetic strength, as well as opposite onsite potentials on the two outer layers.

We first tune the spin-orbit coupling strength of the lower outer layer by replacing the third-layer Hamiltonian with
\begin{equation}
H_3^{(\alpha_{\mathrm{so}})}(\mathbf{k})
=
h_t(\mathbf{k})
-
\alpha_{\mathrm{so}}g(\mathbf{k})
+
h_{\mathrm{m}}(\mathbf{k}),
\label{eq:H3_alpha}
\end{equation}
where
\(
\alpha_{\mathrm{so}}
=
t_{\mathrm{so}}^{(3)}/t_{\mathrm{so}}
\)
measures the spin-orbit-coupling asymmetry between the two outer layers, with $\alpha_{\mathrm{so}}=1$ corresponding to the symmetric case.

We next introduce unequal interlayer couplings by taking
\begin{equation}
t_{12}=t_{\perp},
\qquad
t_{23}=\alpha_t t_{\perp},
\end{equation}
where $t_{12}$ and $t_{23}$ denote the hopping amplitudes between layers $1$ and $2$, and between layers $2$ and $3$, respectively.
The ratio
\(
\alpha_t=t_{23}/t_{12}
\)
measures the interlayer-coupling asymmetry, with $\alpha_t=1$ corresponding to equal interlayer couplings.

We further allow different altermagnetic strengths on the two outer layers by taking
\begin{equation}
J_1=J,
\qquad
J_3=\alpha_J J,
\end{equation}
where $J_1$ and $J_3$ denote the altermagnetic strengths on the upper and lower outer layers, respectively.
The ratio
\(
\alpha_J=J_3/J_1
\)
measures the altermagnetic-strength asymmetry, with $\alpha_J=1$ corresponding to equal strengths.

Finally, we introduce opposite onsite potentials on the two outer layers,
\begin{equation}
H_V
=
V
\begin{pmatrix}
1&0&0\\
0&0&0\\
0&0&-1
\end{pmatrix}
\otimes\tau_0s_0,
\end{equation}
where $V$ denotes the magnitude of the layer-dependent onsite potential.
Accordingly, the upper and lower outer layers acquire onsite shifts $+V$ and $-V$, respectively, while the middle layer remains unchanged.
The fully symmetric model corresponds to
\(
\alpha_{\mathrm{so}}=\alpha_t=\alpha_J=1
\)
and
\(
V=0.
\)

The corresponding results are summarized in Fig.~\ref{fig4}.
Figures~\ref{fig4}(a)--\ref{fig4}(c) show that the four zero-energy states persist over broad ranges of $\alpha_{\mathrm{so}}$, $\alpha_t$, and $\alpha_J$.
For representative asymmetric cases with the corresponding parameter set to $0.8$, the zero-energy states remain strongly localized at the four corners, as shown in Fig.~\ref{fig4}(e).
Thus, moderate asymmetries in the outer-layer spin-orbit coupling, interlayer coupling, and altermagnetic strength do not destroy the corner states.

The opposite outer-layer onsite potential produces a qualitatively different behavior.
As shown in Fig.~\ref{fig4}(d), increasing $V$ drives the four corner-derived levels toward the surrounding low-energy spectrum.
For $V=0.2$, the four states closest to zero energy become spatially extended rather than corner localized, as shown in Fig.~\ref{fig4}(f).
Thus, the opposite layer-dependent onsite potential destroys the corner localization even though low-energy states remain near zero energy.

To distinguish the robustness of the bulk Dirac point from that of the corner states, we further calculate the bulk spectra for representative perturbation strengths.
As shown in Figs.~\ref{fig4}(g)--\ref{fig4}(j), the Dirac crossing at $M$ remains gapless for $\alpha_{\mathrm{so}}=0.8$, $\alpha_t=0.8$, $\alpha_J=0.8$, and $V=0.2$, respectively.
Therefore, all four perturbations considered here preserve the bulk Dirac crossing over the representative parameter regime, whereas their effects on the corner states are different.
In particular, the first three asymmetries preserve both the bulk Dirac point and the corner localization, while the opposite outer-layer onsite potential leaves the bulk Dirac point intact but destroys the corner localization.

We further examine the stability of the corner states against random Zeeman-type onsite disorder,
\begin{equation}
V_{\mathrm{dis}}
=
\sum_i w_i I_{\ell}\tau_0s_z,
\end{equation}
where $I_{\ell}$ is the identity matrix in layer space and $w_i$ is randomly distributed in $[-W/2,W/2]$.
As shown in Fig.~\ref{fig4}(k), the corner states remain pinned near zero energy under moderate magnetic disorder.

These results demonstrate that the bulk Dirac point and the higher-order corner states have distinct robustness conditions.
The coexistence of the two is stable against moderate spin-orbit-coupling, interlayer-coupling, and altermagnetic-strength asymmetries, as well as onsite magnetic disorder, and is therefore not a fine-tuned consequence of the exact layer-space decomposition.
At the same time, the opposite layer-dependent onsite potential provides an example in which the corner localization is destroyed while the bulk Dirac point remains gapless, highlighting the distinct symmetry and sector structures governing the two types of states.

\section{Discussion and conclusion}

We have proposed a route to a two-dimensional higher-order topological Dirac semimetal based on a trilayer nonsymmorphic Dirac system.
Through exact layer-space decomposition, the Hamiltonian separates into a gapless Dirac block and a gapped topological block.
Outer-layer $d$-wave altermagnetism gaps the Dirac crossings at $X$ and $Y$ while preserving the symmetry-protected crossing at $M$, and simultaneously induces opposite Dirac masses on adjacent edges that bind corner states.
This higher-order topological Dirac semimetal phase remains stable against deviations from exact layer-space diagonalization and onsite magnetic disorder.
Our results establish altermagnetism as an effective mechanism for engineering higher-order topology in two-dimensional nonsymmorphic Dirac semimetals.

Finally, from a materials perspective, our proposal should be viewed as a layer engineering principle rather than a material specific prediction.
Nonsymmorphic Dirac bands have been established or proposed in square net and related two dimensional platforms, including ZrSiS, PtPb$_4$, chemically modified group VA monolayers, and surface supported Bi lattices~\cite{Schoop2016,Jin2019,Wang2019Hourglass,Wu2022PtPb4}.
On the magnetic side, layered $d$ wave altermagnets, such as KV$_2$Se$_2$O and Rb$_{1-\delta}$V$_2$Te$_2$O, provide relevant building blocks for momentum dependent exchange fields~\cite{Jiang2025KVSeO,Zhang2025RbVTeO}.
The recently proposed altermagnetic proximity effect further suggests a possible route for imprinting such momentum alternating spin splitting onto adjacent nonmagnetic layers~\cite{Zhu2026AMPE}.
These developments point to a realistic route for combining nonsymmorphic Dirac bands and altermagnetic exchange fields through controlled layer hybridization.

\section*{Acknowledgments}

This work was financially supported by the National Key R and D Program of China (Grant No. 2024YFA1409002), the National Natural Science Foundation of China (Grants No. 12074097, No.  12374034, No.  12547169, and No. 12447147), Natural Science Foundation of Hebei Province (Grant No. A2024205025), the Quantum Science and Technology-National Science and Technology Major Project (Grant No. 2021ZD0302403), and the China Postdoctoral
Science Foundation (Grant No. 2024M760070).
We also acknowledge the High-performance Computing Platform of Peking University for providing computational resources.

\appendix

\section{Exact layer-space block decomposition}
\label{app:block_decomposition}

Here we derive the exact layer-space block decomposition of the trilayer Hamiltonian.
Using the definitions of $h_t(\mathbf{k})$, $g(\mathbf{k})$, and
$h_{\mathrm{m}}(\mathbf{k})$ introduced in the main text, the three layer Hamiltonians are
\begin{align}
H_1(\mathbf{k})
&=
h_t(\mathbf{k})-g(\mathbf{k})+h_{\mathrm{m}}(\mathbf{k}),
\nonumber\\
H_2(\mathbf{k})
&=
h_t(\mathbf{k})+g(\mathbf{k}),
\nonumber\\
H_3(\mathbf{k})
&=
h_t(\mathbf{k})-g(\mathbf{k})+h_{\mathrm{m}}(\mathbf{k}).
\end{align}
The trilayer Bloch Hamiltonian is
\begin{equation}
H_{3\mathrm{L}}(\mathbf{k})
=
\begin{pmatrix}
H_1(\mathbf{k}) & T_{\perp} & 0 \\
T_{\perp} & H_2(\mathbf{k}) & T_{\perp} \\
0 & T_{\perp} & H_3(\mathbf{k})
\end{pmatrix},
\quad
T_{\perp}=t_{\perp}\tau_0s_0.
\label{eq:app_H3L}
\end{equation}

Since the two outer layers are identical,
$H_1(\mathbf{k})=H_3(\mathbf{k})$, we introduce the antisymmetric and symmetric outer-layer states
\begin{equation}
|a\rangle
=
\frac{|1\rangle-|3\rangle}{\sqrt{2}},
\qquad
|s\rangle
=
\frac{|1\rangle+|3\rangle}{\sqrt{2}}.
\end{equation}
In the basis $(|a\rangle,|2\rangle,|s\rangle)$, the layer-space unitary transformation is
\begin{equation}
U
=
\begin{pmatrix}
\frac{1}{\sqrt{2}} & 0 & \frac{1}{\sqrt{2}} \\
0 & 1 & 0 \\
-\frac{1}{\sqrt{2}} & 0 & \frac{1}{\sqrt{2}}
\end{pmatrix},
\end{equation}
where each matrix element is implicitly multiplied by the internal identity matrix $\tau_0s_0$.

Applying this transformation gives
\begin{align}
\widetilde{H}_{3\mathrm{L}}(\mathbf{k})
&=
U^{\dagger}H_{3\mathrm{L}}(\mathbf{k})U
\nonumber\\
&=
\begin{pmatrix}
H_1(\mathbf{k}) & 0 & 0 \\
0 & H_2(\mathbf{k}) & \sqrt{2}T_{\perp} \\
0 & \sqrt{2}T_{\perp} & H_1(\mathbf{k})
\end{pmatrix}.
\label{eq:app_transformed_H3L}
\end{align}
The vanishing matrix element between $|a\rangle$ and $|2\rangle$ follows from
\begin{equation}
\langle a|H_{3\mathrm{L}}|2\rangle
=
\frac{1}{\sqrt{2}}
\left(
T_{\perp}-T_{\perp}
\right)
=
0.
\label{eq:app_cancellation}
\end{equation}
The antisymmetric outer-layer channel is therefore exactly decoupled from the middle layer.

Equation~\eqref{eq:app_transformed_H3L} can be written in the block-diagonal form
\begin{equation}
\widetilde{H}_{3\mathrm{L}}(\mathbf{k})
=
H_a(\mathbf{k})\oplus H_{2s}(\mathbf{k}),
\end{equation}
where
\begin{equation}
H_a(\mathbf{k})
=
h_t(\mathbf{k})-g(\mathbf{k})+h_{\mathrm{m}}(\mathbf{k}),
\end{equation}
and
\begin{equation}
H_{2s}(\mathbf{k})
=
\begin{pmatrix}
h_t(\mathbf{k})+g(\mathbf{k})
&
\sqrt{2}t_{\perp}\tau_0s_0
\\
\sqrt{2}t_{\perp}\tau_0s_0
&
h_t(\mathbf{k})-g(\mathbf{k})+h_{\mathrm{m}}(\mathbf{k})
\end{pmatrix}.
\end{equation}
Introducing Pauli matrices $\eta_i$ in the $(|2\rangle,|s\rangle)$ subspace, the hybridized block becomes
\begin{align}
H_{2s}(\mathbf{k})
={}&
h_t(\mathbf{k})\eta_0
+
g(\mathbf{k})\eta_z
+
\sqrt{2}t_{\perp}\tau_0s_0\eta_x
\nonumber\\
&+
\frac{1}{2}h_{\mathrm{m}}(\mathbf{k})
(\eta_0-\eta_z).
\label{eq:app_H2s_block}
\end{align}

The decomposition is exact whenever the two outer layers are identical and their couplings to the middle layer are equal.
It therefore remains valid at finite $J$ as long as the altermagnetic term is identical on the two outer layers.

\section{Altermagnetic effect on the Dirac sector} 
\label{sec:finiteJ_Dirac} 

At finite $J$, the two outer layers remain identical, and hence the exact layer-space decomposition remains valid. We first analyze the effect of the altermagnetic term on the antisymmetric Dirac block $H_a$.

The antisymmetric block at finite $J$ is
\begin{equation}
H_a(\mathbf{k})
=
h_t(\mathbf{k})
-
g(\mathbf{k})
+
h_{\mathrm{m}}(\mathbf{k}).
\label{eq:app_Ha_finiteJ}
\end{equation}
At $J=0$, this block hosts Dirac points at $X$, $Y$, and $M$.
The scalar form factor of the altermagnetic term,
\begin{equation}
f_d(\mathbf{k})=\cos k_x-\cos k_y,
\end{equation}
satisfies
\begin{equation}
f_d(X)=-f_d(Y),
\qquad
f_d(M)=0,
\label{eq:app_fd_TRIM}
\end{equation}
because it changes sign under the fourfold rotation $C_{4z}$.

To make the resulting mass structure explicit, we expand the Hamiltonian around the three Dirac points.
For $\mathbf{k}=X+\mathbf{q}$, $\mathbf{k}=Y+\mathbf{q}$, and $\mathbf{k}=M+\mathbf{q}$, respectively, the leading terms are
\begin{align}
H_a^{X}(\mathbf{q})
\simeq{}&
-tq_x\tau_xs_0
+
t_{\mathrm{so}}\tau_z
(q_xs_y+q_ys_x)
-
2J\tau_zs_z,
\label{eq:app_HaX}
\\
H_a^{Y}(\mathbf{q})
\simeq{}&
-tq_y\tau_xs_0
-
t_{\mathrm{so}}\tau_z
(q_xs_y+q_ys_x)
+
2J\tau_zs_z,
\label{eq:app_HaY}
\\
H_a^{M}(\mathbf{q})
\simeq{}&
t_{\mathrm{so}}\tau_z
(q_xs_y-q_ys_x)
+
\frac{t}{2}q_xq_y\tau_xs_0
\nonumber\\
&+
\frac{J}{2}(q_x^2-q_y^2)\tau_zs_z.
\label{eq:app_HaM}
\end{align}

The momentum independent terms in Eqs.~\eqref{eq:app_HaX} and \eqref{eq:app_HaY} anticommute with the corresponding linear Dirac Hamiltonians and therefore generate masses
\begin{equation}
m_X=-2J\tau_zs_z,
\qquad
m_Y=+2J\tau_zs_z.
\label{eq:app_mass_XY}
\end{equation}
The opposite signs of $m_X$ and $m_Y$ directly reflect the sign changing $d$ wave form factor.

At $M$, by contrast, the altermagnetic contribution begins at quadratic order in $\mathbf{q}$ and vanishes exactly at the Dirac point.
It therefore cannot generate a constant mass for the linear Dirac Hamiltonian.
Consequently, finite $J$ gaps the Dirac crossings at $X$ and $Y$, while the Dirac crossing at $M$ remains gapless.

\section{Boundary mass and corner states}
 \label{sec:finiteJ_boundary}

We next consider the hybridized block $H_{2s}$.
Expanding Eq.~\eqref{eq:app_H2s_block} around $\Gamma$ to second order in momentum gives
\begin{align}
H_{2s}(\mathbf{k})
\simeq{}&
\left[
2t-\frac{t}{4}(k_x^2+k_y^2)
\right]\tau_xs_0\eta_0
+
t_{\mathrm{so}}\tau_z
(k_xs_y-k_ys_x)\eta_z
\nonumber\\
&
+
\sqrt{2}t_{\perp}\tau_0s_0\eta_x
-
\frac{J}{4}(k_x^2-k_y^2)\Gamma_m,
\label{eq:app_H2s_cont}
\end{align}
where
$
\Gamma_m
=
\tau_zs_z(\eta_0-\eta_z).
$
The first three terms describe the parent quantum spin Hall block, whereas the last term is the continuum altermagnetic perturbation.
Its $k_x^2-k_y^2$ dependence is responsible for the opposite masses on orthogonal boundaries.

Consider edge $I$ in Fig.~\ref{fig3}(b), described by the semi-infinite geometry $x\geq0$, with $k_y\equiv k_{\parallel}$ conserved.
After replacing $k_x\rightarrow-i\partial_x$ and neglecting terms of order $k_{\parallel}^2$, Eq.~\eqref{eq:app_H2s_cont} can be written as
\begin{align}
H_0^{I}(-i\partial_x)
={}&
-i t_{\mathrm{so}}\partial_x
\tau_zs_y\eta_z
+
\left(
2t+\frac{t}{4}\partial_x^2
\right)\tau_xs_0\eta_0
\nonumber\\
&+
\sqrt{2}t_{\perp}\tau_0s_0\eta_x,
\label{eq:app_H0I}
\\
H_p^{I}(-i\partial_x,k_{\parallel})
={}&
-t_{\mathrm{so}}k_{\parallel}
\tau_zs_x\eta_z
+
\frac{J}{4}\partial_x^2\Gamma_m.
\label{eq:app_HpI}
\end{align}

Choosing the parent edge Dirac point as zero energy, the two boundary states satisfy
\begin{equation}
H_0^{I}(-i\partial_x)\psi_{\alpha}^{I}(x)=0,
\qquad
\psi_{\alpha}^{I}(0)
=
\psi_{\alpha}^{I}(\infty)
=
0,
\label{eq:app_edge_zero}
\end{equation}
where $\alpha=1,2$ labels the Kramers pair of the parent quantum spin Hall block.

Projecting $H_p^{I}$ onto this helical edge subspace gives
\begin{equation}
\left[
H_I(k_{\parallel})
\right]_{\alpha\beta}
=
\int_0^{\infty}dx\,
\left[
\psi_\alpha^{I}(x)
\right]^{\dagger}
H_p^{I}
\psi_\beta^{I}(x).
\label{eq:app_projection}
\end{equation}
Up to an overall energy shift and a unitary rotation within the edge subspace, the resulting Hamiltonian takes the massive Dirac form
\begin{equation}
H_I(k_{\parallel})
=
v_Ik_{\parallel}\mu_z
+
M_I\mu_y,
\label{eq:app_edge_HI}
\end{equation}
where $\mu_i$ are Pauli matrices acting on the two helical edge states.
The mass coefficient $M_I$ is determined by the projected altermagnetic perturbation,
\begin{equation}
M_I
=
\frac{J}{4}
\mathcal{M}_I,
\qquad
\mathcal{M}_I
=
\int_0^{\infty}dx\,
\Psi_I^{\dagger}(x)
\partial_x^2\Gamma_m
\Psi_I(x),
\label{eq:app_mass_I}
\end{equation}
where $\Psi_I=(\psi_1^{I},\psi_2^{I})$ denotes the edge state doublet and the matrix element is understood to be expressed in the basis where the mass term is proportional to $\mu_y$.
This symmetry allowed matrix element is nonzero for the present boundary states, consistent with the finite edge gap in the ribbon spectrum.

The mass relation between orthogonal edges follows directly from symmetry.
At $J=0$, the edge subspaces on two orthogonal boundaries are related by $C_{4z}$.
Meanwhile, the altermagnetic form factor satisfies
\begin{equation}
f_d(C_{4z}\mathbf{k})
=
-f_d(\mathbf{k}).
\label{eq:app_fd_C4}
\end{equation}
The projected mass therefore changes sign under a ninety degree rotation,
\begin{equation}
M_{II}=-M_I.
\label{eq:app_mass_orthogonal}
\end{equation}
Opposite parallel edges are related by the twofold rotation and carry the same mass sign.
For the four successive edges $I$, $II$, $III$, and $IV$ of the square geometry,
\begin{equation}
M_{II}=-M_I,
\qquad
M_{III}=M_I,
\qquad
M_{IV}=-M_I.
\label{eq:app_mass_pattern}
\end{equation}
Adjacent edges therefore carry Dirac masses with opposite signs.
Each corner is consequently a mass domain wall of the one-dimensional edge theory and binds a localized zero-dimensional state through the Jackiw-Rebbi mechanism~\cite{JackiwRebbi1976}.

\section{Mirror-graded winding number of the hybridized sector}
\label{app:mirror_winding}

We further characterize the higher-order topology of the gapped hybridized sector $H_{2s}$.
At finite $J$, this block is
\begin{align}
H_{2s}(\mathbf{k})
={}&
2t\cos\frac{k_x}{2}\cos\frac{k_y}{2}\,
\tau_x s_0\eta_0
\nonumber\\
&+
t_{\mathrm{so}}\tau_z
\left(
s_y\sin k_x-s_x\sin k_y
\right)\eta_z
+
\sqrt{2}t_{\perp}\tau_0s_0\eta_x
\nonumber\\
&+
\frac{J}{2}
(\cos k_x-\cos k_y)
\tau_zs_z(\eta_0-\eta_z).
\label{eq:app_H2s_mirror}
\end{align}
Although the full trilayer system is gapless because of the Dirac block $H_a$, the block $H_{2s}$ remains fully gapped and can therefore be characterized independently.

We first consider the diagonal mirror-invariant line
$
k_x=-k_y\equiv k.
$
Since $\cos k_x-\cos k_y=0$ along this line, Eq.~\eqref{eq:app_H2s_mirror} becomes
\begin{align}
H_{2s}(k,-k)
=&
t(1+\cos k)\tau_xs_0\eta_0
+
\sqrt{2}t_{\mathrm{so}}\sin k\,
\tau_zs_{+}\eta_z
\nonumber\\
&+
\sqrt{2}t_{\perp}\tau_0s_0\eta_x,
\end{align}
where
\begin{equation}
s_{+}=\frac{s_x+s_y}{\sqrt{2}}.
\end{equation}
Because $[H_{2s}(k,-k),s_{+}]=0$, the Hamiltonian decomposes into two mirror sectors $m=\pm1$,
\begin{equation}
H_m^{(-)}(k)
=
t(1+\cos k)\tau_x\eta_0
+
m\sqrt{2}t_{\mathrm{so}}\sin k\,
\tau_z\eta_z
+
\sqrt{2}t_{\perp}\tau_0\eta_x.
\end{equation}
Each mirror block possesses the chiral symmetry
\begin{equation}
\{\Gamma,H_m^{(-)}(k)\}=0,
\qquad
\Gamma=\tau_z\eta_y,
\end{equation}
and commutes with
$
Q=\tau_x\eta_x.
$
Since $[\Gamma,Q]=0$, each mirror sector can be further decomposed into $Q=r=\pm1$ subsectors.
The resulting two-band Hamiltonian is
\begin{equation}
H_{m,r}^{(-)}(k)
=
\left[
t(1+\cos k)+r\sqrt{2}t_{\perp}
\right]\rho_z
+
m\sqrt{2}t_{\mathrm{so}}\sin k\,\rho_x.
\end{equation}
In the chiral basis, the corresponding off-diagonal function is
\begin{equation}
q_{m,r}^{(-)}(k)
=
t(1+\cos k)
+
r\sqrt{2}t_{\perp}
+
im\sqrt{2}t_{\mathrm{so}}\sin k,
\end{equation}
which defines the winding number
\begin{equation}
\nu_{m,r}^{(-)}
=
\frac{1}{2\pi i}
\int_{-\pi}^{\pi}
dk\,
\partial_k
\ln q_{m,r}^{(-)}(k).
\end{equation}
A nonzero winding occurs when
\begin{equation}
\left|
t+r\sqrt{2}t_{\perp}
\right|
<
|t|.
\end{equation}
For $t=-1$ and $t_{\perp}=0.3$, only the $r=+1$ subsector is nontrivial.
The two mirror sectors carry opposite winding numbers, and the corresponding mirror-graded winding number,
\begin{equation}
\nu_{\mathcal{M}_{-}}
=
\frac{\nu_{+,+}^{(-)}-\nu_{-,+}^{(-)}}{2},
\end{equation}
satisfies
$
|\nu_{\mathcal{M}_{-}}|=1.
$

The same analysis applies to the other diagonal mirror-invariant line $k_x=k_y\equiv k$.
In this case, the conserved mirror operator is
\begin{equation}
s_{-}=\frac{s_y-s_x}{\sqrt{2}},
\end{equation}
and the resulting mirror-sector Hamiltonians have the same reduced form as above.
Consequently, the corresponding mirror-graded winding number also satisfies
\begin{equation}
|\nu_{\mathcal{M}_{+}}|=1.
\end{equation}
Therefore,
$
|\nu_{\mathcal{M}_{+}}|
=
|\nu_{\mathcal{M}_{-}}|
=
1,
$
providing an independent bulk characterization of the higher-order topology of the gapped $H_{2s}$ sector.
These nonzero invariants are consistent with the opposite effective Dirac masses on adjacent edges derived in Appendix~\ref{sec:finiteJ_boundary} and with the corner-localized states in the finite nanoflake.

\end{document}